\documentclass[prl,aps,twocolumn]{revtex4}

\usepackage[mathscr]{eucal}
\usepackage[latin1]{inputenc}
\usepackage{amsmath}
\usepackage{amsfonts}
\usepackage{amssymb}
\usepackage{pictex}
\usepackage{graphicx}

\def\be{\nopagebreak[3]\begin{equation}}
\def\ee{\end{equation}}
\def\ba{\nopagebreak[3]\begin{eqnarray}}
\def\ea{\end{eqnarray}}

\def\d{{\rm d}}

\newcommand{\teta}{\rlap{\lower2ex\hbox{$\,\tilde{}$}}\eta{}}

\def\lp{{\ell}_{\rm Pl}}

\def\bb{{\tt b}}

\def\co{\sqrt{12 \pi G}}

\newcommand{\f}{\frac}

\newcommand{\ep}{\epsilon}

\usepackage{enumerate}

\usepackage{colordvi}
\newcounter{mnotecount}[section]

\newcommand{\comment}[1]{}

\def\f{\frac}

\def\co{\beta}

\def\bet{\beta}

\begin{document}
\title{Quantum bounce and cosmic recall}
\author{Alejandro Corichi$^{\dagger,\star}$ and
Parampreet Singh$^{\ddagger,\star}$}
\affiliation{${}^\dagger$Instituto de Matem\'aticas,
Unidad Morelia, Universidad Nacional Aut\'onoma de
M\'exico, UNAM-Campus Morelia, A. Postal 61-3, Morelia, Michoac\'an 58090,
Mexico \\ ${}^\star$Center for Fundamental Theory, Institute for Gravitation
and the Cosmos, Penn State,
University Park
PA 16802, USA \\ ${}^\ddagger$ Perimeter Institute for Theoretical Physics,
31 Caroline Street North, Waterloo, Ontario N2L 2Y5, Canada}

\begin{abstract}
Loop quantum cosmology predicts that, in simple models, the big
bang singularity of  classical general relativity is replaced by a
\emph{quantum bounce}. Because of the extreme physical conditions
near the bounce, a natural question is whether the universe can
retain, after the bounce, its memory about the previous epoch.
More precisely, does the universe retain 
various properties of the
state after evolving unitarily through the bounce or does it
suffer from some cosmic amnesia as has been recently suggested? Here we
show that this issue can be answered unambiguously at least within an
exactly solvable model, derived from a small simplification of
loop quantum cosmology,  for which there is full analytical
control on the quantum theory. We show that if there exists a
semiclassical state at late times on one side, peaked around a
pair of canonically conjugate variables, then there are very
strong bounds on the fluctuations on the other side of the bounce,
implying semi-classicality. For such a model universe which grows to a
size of 1 megaparsec,
the change in relative fluctuations of the only non-trivial
observable of the model
across the bounce is less than $10^{-56}$ (becoming smaller for
universes which grow larger). The universe maintains (an almost)
total recall.

\end{abstract}

\maketitle


The singularity theorems of classical general relativity (GR) have
taught us that the homogeneous cosmological solutions possessing a
singularity are examples of a generic feature of the theory,
rather than an artifact of the symmetries they exhibit
\cite{sing}. In quantum gravity we do not yet have a general
result on the fate of the classical singularities. The detailed
study of isotropic solutions in a symmetry reduced quantum theory
of gravity known as loop quantum cosmology (LQC) \cite{lqc} has
shown that for a universe filled with a massless scalar field,
with or without a cosmological constant, the singularity gets
replaced by a \emph{quantum bounce} \cite{APS}. Whether
singularity resolution is an artifact of the symmetries or a
generic feature of  quantum cosmology is an open question. However, homogeneous models
are excellent approximation to our Universe at large scales and recent analytical
and numerical studies in GR strongly suggest that homogeneous modes dictate the 
dynamics near space-like singularities (thus confirming the Belinskii-Khalatnikov-Lifshitz conjecture). Investigations of simple models as above, even with their intrinsic
limitations, thus provide valuable insights
on difficult questions such as the nature of spacetime across 
classical singularities. Just as in other instances in physics,
one could expect that qualitative features of these models survive in more 
general situations. Preliminary studies of anisotropic models within LQC
suggest that such expectations are justified
\cite{chiou}.

Within this model, extensive analytical and numerical investigations 
have revealed various features of the quantum spacetime
across the bounce \cite{APS,APSV,warsaw}. Due to the complicated
form of the quantum constraint,  full analytical control of LQC is
difficult, so it is natural to look for slightly simpler models
that capture the physical content of the theory but that are
amenable for analytical investigations. Recently, such a model
based on a small simplification of the quantum constraint in $k$=0
isotropic LQC has been introduced \cite{ACS}. It is exactly
solvable and replicates all the features of LQC for
universes which grow to a large size to a great accuracy. This model describes
homogeneous and isotropic flat universe sourced with a massless
scalar field $\phi$ (which plays the role of internal clock). The
exactly solvable model of LQC (sLQC) not only confirms the result
of the quantum bounce for semi-classical states as first seen in the
numerical simulations \cite{APS} but extends them to almost all
the states of the physical Hilbert space. Further, the quantum
constraint approximates the classical one at both early and late
times. Various questions which could not be fully answered in LQC,
can now be posed in sLQC. 

Since the curvatures are so large at the bounce, it is natural to
inquire whether significant `information' can be lost across it.
Such a possibility was recently highlighted in \cite{nature} where
it was argued that a universe such as ours, which is
semi-classical to an excellent degree of approximation, could have
nonetheless evolved from a state which has an arbitrarily large
(and unconstrained) relative dispersions 
at early times
much before the bounce. If this were possible, by making
observations today or in the future we would never be able to
determine  the initial, pre-bounce state of the universe. 
The purpose of this Letter is to address this question within sLQC.

Our analysis is based on this simplified model that features, however, 
two significant improvements over that of \cite{nature,MB,mb2}.
First, it is very closely related to LQC through a simple, well
motivated and well controlled approximation of the quantum
constraint. Secondly,
sLQC 
agrees with general relativity at low curvatures and is
free from gauge artifacts. Further, our analysis is not tied to
the coherence properties of the state which is a stronger
assumption than semi-classicality which only requires for an
observable ${\cal O}$ that its expectation value be very close to the
classical one and
the relative fluctuation $(\Delta {\cal \hat O})^2/\langle {\cal
\hat O} \rangle^2 \ll 1$. We will show that the relative
fluctuations in the initial semi-classical state peaked in
conjugate variables, strongly bound the relative fluctuations on
the other side of the bounce. More precisely the relative
fluctuation of the Dirac observable $\hat V|_\phi$ (volume at a
given instant of `time' $\phi = \phi_o$) across the bounce
satisfies the above inequality if the initial state was peaked in
a large classical universe. In fact, for a large class of states
the change in relative dispersion is zero. 
Since the field $\phi$ is massless, the expectation value of its
momentum $\hat p_\phi$ and corresponding fluctuations
characterizing  the state are trivially conserved across the
bounce. Thus, above results show that a state which is
semi-classical at late (early) times, preserves semi-classicality
at early (late) times across the bounce.

Let us recall the model we are considering. In terms
of the phase space variables used in loop quantum gravity (LQG)
\cite{lqg}, namely a connection $A^i_a$ and a densitized triad
$E^a_i$, the homogeneous sector can be expressed as,
\be A_a^i=c\,V_0^{-(1/3)}\,{}^{\rm o}\omega^i_a\quad ;
\quad E^a_i=p\, V_0^{-(2/3)}\,\sqrt{q_0}\,{}^{\rm o}e_i^a
\ee
where $({}^{\rm o}\omega^i_a,{}^{\rm o}e_i^a)$ are a set of
orthonormal co-triads and triads compatible with the fiducial
(flat for $k=0$) metric ${}^{\rm o}q_{ab}$ and $V_o$ is the volume of the fiducial cell, introduced to define sympectic structure, with respect to  ${}^{\rm o}q_{ab}$.
The  phase space is
characterized by conjugate variables $(c,p)$ satisfying 
$\{c,p\}=(8\pi G\gamma)/3$.
%
where $\gamma$ 
is the Barbero-Immirzi parameter. 
 Triad $p$ is
related to the physical volume $V$ as $|p| = V^{2/3} =
V_o^{2/3} a^2$ where $a$ is the scale factor and on the space of classical
solutions $c = \gamma \dot a$.
It is convenient to introduce \cite{ACS}
\be
{\tt b}:=\varepsilon\,c/p^{1/2}\qquad{\rm and}\qquad
\nu = \varepsilon\,p^{3/2}/(2 \pi \lp^2 \gamma) ~ \label{b_eq}
\ee
where $\hbar\{\bb,\nu\} = 2$. Here $\varepsilon = \pm 1$ is the
orientation of the triad with respect to that of ${}^{\rm
o}\omega^i_a$. The classical constraint then becomes
$\bb^2 V^2/\gamma^2 = (8 \pi G/3) p_\phi^2/2$.
 In the $\bb$ representation, the quantum constraint of sLQC becomes \cite{ACS},
\be
\f{\partial^2}{\partial \phi^2}\cdot\chi(\bb,\phi)=
\bet^2 \,
\left(\frac{\sin(\lambda\bb)}{\lambda}\;\frac{\partial}{\partial\,
\bb} \right)^2
\cdot \chi(\bb,\phi)\label{const1}
\ee
with $\bb\in (0,\pi/\lambda)$, $\bet := \sqrt{12 \pi G}$ and
$\lambda^2 = 2 \sqrt{3} \pi \gamma \lp^2$ \cite{ACS}.
 By introducing
\be
x = \bet^{-1} \ln (\tan(\lambda \bb/2)) \label{x-eq}
\ee
the quantum constraint can be rewritten in a Klein-Gordon form
\be
\partial_\phi^2 \, \chi(x,\phi) = \partial_x^2 \, \chi(x,\phi) ~.
\ee

A general solution $\chi(x,\phi)$ to the above equation can be
decomposed in the left and right moving components:
\be\label{chi_eq} \chi = \chi_+ (\phi + x) + \chi_-(\phi - x) :=
\chi_+(x_+) + \chi_-(x_-) ~.
\ee
The physical states are positive frequency solutions of
($\ref{chi_eq}$). Since there are no fermions in the model, the
orientations of the triad are indistinguishable and $\chi(x,\phi)$
satisfy the symmetry requirement $\chi(-x,\phi) = -\chi(-x,\phi)$.
Thus,
$\chi(x,\phi) =  (F(x_+) - F(x_-))/\sqrt{2}$, where $F$ is an
arbitrary positive frequency solution.
 The physical inner product given as \cite{ACS}
\be (\chi_1,\chi_2)_{\mathrm{phy}} = - 2 i \,
\int_{-\infty}^{\infty} \, \d x \, \bar F(x) \, \partial_x F(x) ~.
\ee
We can now compute the expectation values and fluctuations of
$\hat V|_{\phi{_o}}$.  For {\it any} state of the physical Hilbert
space
\begin{eqnarray}
\langle\hat{V}\rangle_\phi &=& V_+\,e^{\, \bet \,\phi}
+V_-\,e^{-\bet\,\phi} \label{v-exp} ~,\\
\langle\hat{V}^2\rangle_\phi~&=& \hskip-0.1cm W_0+W_+\,e^{2\bet\,\phi}
+W_-\,e^{-2\bet\,\phi}\,, \label{fluct} \\
(\Delta\hat{V})_\phi^2&=&Y_0+Y_+\,e^{2\bet\,\phi}
+Y_-\,e^{-2\bet\,\phi} \label{fluct2} ~,
\end{eqnarray}
with $V_{\pm}$,  $W_0$, $W_{\pm}$, $Y_0$ and $Y_\pm$ being  real and positive, given by
\ba
V_{\pm}=\frac{4 \pi \gamma \lp^2
\lambda}{\co}\int\d x \left|\frac{\d F}{\d
x}\right|^2\,e^{\mp\bet\, x} \label{const5} ~,\nonumber \\
%
%
W_0= \f{i \pi \gamma^2\lp^4\lambda^2}{3G} \int\d
x\left(\f{\d^2\bar{F}}{\d
x^2}\,\f{\d F}{\d x} - \f{\d \bar{F}}{\d x}\,\f{\d^2 F}{\d x^2}\right) ~,\nonumber 
\ea
\be
\label{w-eq}
W_\pm=\f{i\pi \gamma^2\lp^4\lambda^2}{6G} \int\d x\;e^{\mp 2\co\,x}
\left(\f{\d^2\bar{F}}{\d
x^2}\,\f{\d F}{\d x} - \f{\d \bar{F}}{\d x}\,\f{\d^2 F}{\d x^2}\right) ~,
\ee
$Y_0 = W_0 - 2 V_+ V_-$ and $Y_\pm = W_\pm - V_\pm^2$.

From (\ref{v-exp}), it follows that the expectation value of $\hat
V|_\phi$ is large at both very early and late times and  has a
non-zero global minimum $V_{\mathrm{min}} = 2 (V_+
V_-)^{1/2}/||\chi||^2$ occurring at bounce time $\phi_b^V = (2
\bet)^{-1} \ln(V_-/V_+)$ \cite{ACS,MB}. Around $\phi = \phi_b^V$, $\langle
\hat V\rangle_\phi$ is symmetric. Similarly, $\langle \hat V^2
\rangle_\phi$ is symmetric across $\phi_b^{V^2}=(4
\bet)^{-1}\ln(W_-/W_+)$. The first resulting observation is that
if $\phi_b^V = \phi_b^{V^2}$, the difference in the asymptotic
values of the relative fluctuation
\be\label{D-eq}
D:=\lim_{\phi\to\infty}\left[\left(\frac{(\Delta\hat{V})}{\langle\hat{V}\rangle}\right)_{-\phi}^2
-
\left(\frac{(\Delta\hat{V})}{\langle\hat{V}\rangle}\right)_{\phi}^2\right] = \frac{W_-}{V^2_-}-\frac{W_+}{V^2_+}
\ee
vanishes.

We now show that for a large class of states, the difference
between the relative fluctuations $D$ is indeed zero. We first
express a physical state in terms of $F$
\be
\chi(x,\phi) = 
2i\int \,\d k
\;\tilde{F}(k)\,e^{-ik\,\phi}\sin(k\,x) ~.
\ee
In the $\nu$ representation:  $\tilde{e}_k(\nu)=\nu\, {\cal F}^{-1}[e_k(x(b))]$,
with $e_k(x)=\sin(k\,x)$ and where ${\cal F}$ denotes the Fourier
transform. Note that $\tilde{e}_k(\nu)$ is imaginary and symmetric
in $\nu$. Hence, if  $\tilde{F}(k)$ is real, then
$\Psi(\nu,\phi)=\overline{\Psi(\nu,-\phi)}$.~
We can now evaluate the fluctuations:
\ba (\Delta \hat V)^2(\phi) &=&
\langle\Psi|(\hat{V}-\bar{V})^2|\Psi\rangle_{\rm phy}
\nonumber\\
&=& \sum_m
|V_m-\bar{V}|^2\,\overline{\Psi(\nu_m,\phi)}\,\Psi(\nu_m,\phi)
\ea
which show that if $\tilde{F}(k)$ is real, then $(\Delta
\hat V)^2_\phi=(\Delta \hat V)^2_{-\phi}$. In fact, it turns out that the
relative fluctuations are also symmetric (given that, by the same
argument, $\langle \hat{V}\rangle(\phi)= \langle
\hat{V}\rangle(-\phi)$). Further, it is straightforward to see
that this argument also extends to the initial state  (in $k$) $
\tilde{F}(k)\,e^{ik\,x_0}$, which in `$x$ representation' is
`peaked' around $x=x_0$ (or is displaced). In this case, the
fluctuations are symmetric with respect to the time $(\phi-x_0)$.
Note that, this class of symmetric states, includes any real
linear combination $\sum_n \alpha_n\,f_n(k)$ of functions of the
form $f_n(k)=k^n\;e^{-(k-k_0)^2/\beta^2+ik\,x_0}$. As examples,
squeezed states with {\it arbitrary} squeezing, and states
having a Poisson distribution belong to this class. All such
states preserve the relative fluctuation across the bounce. (A result that
holds for dynamical coherent states \cite{MB,mb2}).

It is however possible that a realistic model universe is described by a
state which does not belong to the above class and for which $D$
would not vanish. We now consider such states describing at late
times a low curvature, large classical universe. For concreteness
we wish to look at the evolution of fluctuations as we look
backwards through the bounce from the expanding branch. A state
describing the present (and future) late time epoch shall be
sharply peaked on the classical trajectories and thus the relative
dispersion in $\hat V|_\phi$ shall satisfy: \be
\lim_{\phi\to\infty}(\Delta
\hat{V}_\phi/\langle\hat{V}\rangle_\phi)^2=
W_+/V^2_+-1=:C-1=:\delta \ll 1 ~. \ee
that has to  be compared with the
asymptotic value of the relative dispersion `before the big bang':
\be
\lim_{\phi\to-\infty}(\Delta
\hat{V}_\phi/\langle\hat{V}\rangle_\phi)^2= W_-/V^2_--1=:B-1\,
. \ee
The quantity of interest is $D=B-C$, introduced in (\ref{D-eq}).
To evaluate it, let us consider the ratio of $B$ and $C$, $E :=
B/C = (W_-V^2_+)/(W_+V_-^2)$. Let us assume that the relative
fluctuations are larger on the other side of the bounce (the
pre-big branch in the considered case). Then $E$ is greater than
unity and we parameterize it as $E:=(1+\tilde{\Delta})$.
We now show that if the dispersion of  canonically conjugate
variable to $\hat{V}$ and the relative dispersion of $\hat V$ in
the present epoch is small ($\delta\ll1$), then both of the
quantities $\tilde{\Delta}$ and $D$ are small. This  in turn will
imply a full control on the fluctuations of the universe `before
the big bang'.

Let us start by considering, as initial data, wave functions in
$x$ that have most of its support in the interval
$(x_0-\epsilon,x_0+\epsilon)$, around the point $x_0$.
Intuitively, the quantity $\epsilon$ is proportional to spread or
the absolute dispersion $\Delta x$ and using (\ref{x-eq}) to the
relative dispersion $\Delta \tan(\lambda \bb/2)/\tan(\lambda
\bb/2)$. At low curvature scales, $\bb \ll 2/\lambda$, and thus
$\epsilon$ becomes proportional to $\Delta \bb/\bb$, i.e. the
relative dispersion in the conjugate variable to $V$. If the state
has to describe a universe at low curvature scales,  $\epsilon$ is
required to be small.

In order to put bounds on $\tilde \Delta$ we note that there
exists some freedom in the choice of initial state. This freedom
is in the choice of time $\phi$ at which the quantities are
computed. After all, the wave-functions propagate along
characteristics $x=\pm\phi$ preserving its shape (a feature
responsible for invariance of $\langle \hat p_\phi \rangle$ and
$(\Delta \hat p_\phi)^2$). We  use this freedom, and without loss
of generality, go to the time
$\phi=\phi_b^{V^2}$, that is, the time at which
$\langle\hat{V}^2\rangle_\phi$ bounces, where the $V$'s and $W$'s are
computed. With this choice  $W_+=W_-$.

With this choice, the problem is simplified and  we only have to
consider the ratio $1+\tilde{\Delta}=V^2_+/V^2_-$. Using
(\ref{const5}) for the constants $V_\pm$, we can then put bounds
when the state is supported in the interval
$(x_0-\epsilon,x_0+\epsilon)$:
$$
V_+ < N\,e^{-\co\;(x_0-\epsilon)}\quad ;
\quad
V_- > N\,e^{\co\;(x_0-\epsilon)}
$$
with $N=\frac{16 \gamma \lp^2 \lambda^2}{\co}\int\d x
\left|\frac{\d F}{\d x}\right|^2$. We thus obtain, \be
1+\tilde{\Delta}=\f{V^2_+}{V^2_-}<e^{-4\co\;(x_0-\epsilon)} ~. \ee
Unless $x_0$ is bounded, this inequality provides little insights.
For $W_+ = W_-$, consistency of equations demands that
$0\in(x_0-\epsilon,x_0+\epsilon)$. From (\ref{w-eq}) it is easy to
see that if this were not the case, we would have the same
function
weighted with functions that are larger than one
for, say, $W_+$ and smaller than one for $W_-$. The integrals
would thus be different, contradicting
$W_+=W_-$. Hence, $0\in(x_0-\epsilon,x_0+\epsilon)$. Thus,
it follows that
$\epsilon > |x_0|$, from which we have,
$
1+\tilde{\Delta}=\frac{V^2_+}{V^2_-}<e^{8\co\;\epsilon}
$
and therefore, $\tilde{\Delta} < e^{8\co\,\epsilon} - 1$. Finally,
we obtain
\ba
D &=& B-C = (1+\tilde{\Delta})\, C - C = \tilde{\Delta}\,C \nonumber\\
&=& (1+\delta)\,
\tilde{\Delta} < (1+\delta)\,(e^{8\co\;\epsilon} - 1) \label{res2}
\ea
which is arbitrary small when \emph{both} $\delta$ and $\epsilon$
are small. Since, $\delta$ and $\epsilon$ are measures of relative
dispersions in volume $V$ and curvature $\bb$ respectively,  a
state describing a large low curvature universe  indeed have
$\delta$ and $\epsilon$ very small. Thus,  a state which is
sharply in $V|_\phi$ and its conjugate variable (or in $p_\phi$
for solutions to the constraint) on one side of the bounce is also
sharply peaked on the other side. 

We are now in position to answer the primary question of this
Letter. Does this model universe retain its semi-classicality across the
bounce? What we have shown is that for a state that is
semiclassical at late times and for which the fluctuations in the
volume are {\it not} symmetric across the bounce, there is
nevertheless a strong bound on the possible dispersion on the
`other side of the bounce', making again the state semiclassical
`at early times'. This result was
obtained within a simple but completely solvable model
(sLQC) for which there is full control on the quantities at play.
Note that our bounds are stringent only for states which are truly
semi-classical at late times; they do not tell us much about
states which are sharply peaked in volume but not its conjugate
variable.

For the states of interest which describe at late times a large,
low curvature universe both $\beta\,\epsilon$ and $\delta$ are
small and to the leading order the inequality (\ref{res2}) yields $D < 8 \, \bet \, \ep$.
The value of
$\epsilon$ in a classical universe can be heuristically estimated
as follows. We first note that for sharply peaked states using
$(\ref{x-eq})$, it follows that $\bet \, \ep \sim \bet \Delta x
\sim (\Delta \tan(\lambda \bb/2))/\tan (\lambda \bb/2)$. At low
curvatures, $\tan(\lambda \bb/2) \approx \lambda \bb/2$ and hence
$\bet \, \ep \sim (\Delta \bb)/\bb$. Introducing a parameter $\alpha_1$ which measures the departure
from minimum uncertainty between $\bb$ and $\nu$, such that
$\Delta \bb \, \Delta \nu \approx 2 \alpha_1$, for the state
peaked on the above classical trajectory 
\be
(\Delta \bb/\bb) \, (\Delta \nu/\nu) \approx \sqrt{12 \pi/G} \, \, \alpha_1 \lp^2/p_\phi
\ee
which leads to (in classical $c = G = 1$
units) \be\label{rel-disp-b} (\Delta \bb/\bb)^2 \approx \sqrt{12
\pi} \, \alpha_1 \alpha_2 (\hbar/p_\phi) \ee
where $\alpha_2$ is a parameter which measures the ratio of
relative dispersions in the conjugate variables:
$\alpha_2 := (\Delta \bb/\bb)/(\Delta \nu/\nu)$. 

In our units, $p_\phi$ has dimensions of $\hbar$, thus for a
universe  to have a classical interpretation: $p_\phi \gg \hbar$
(implying $\Delta \bb/\bb \ll 1$). In order to estimate $p_\phi$
it is useful to consider a closed universe which grows to a
maximum size of 1 megaparsec. Using Friedman equation, the value
of   $p_\phi$ for the scalar field filling such a universe is
$p_\phi \sim 1.8 \times 10^{114} \hbar$. Thus, for the initial
state peaked at a classical trajectory in a 1 megaparsec universe:
 $\bet \, \ep \sim (\Delta \bb/\bb) \approx 2 \, \sqrt{\alpha_1 \alpha_2} \, \times 10^{-57}$. Taking $(\alpha_1,\alpha_2) \sim O(1)$ for the initial semi-classical state, we obtain $D < 10^{-56}$, which is an
extremely small number. Thus across the bounce, $(\Delta \hat
V_\phi/\langle \hat V \rangle_\phi)^2 \ll 1 $ and the state
retains semi-classicality.
A similar calculation for a universe with a size resembling ours,
yields the change in relative fluctuations $D < 10^{-60}$. Hence,
within this  model, the variation in the relative dispersion for 
a `realistic universe' across the bounce is negligible. 
The universe recalls the properties of the semi-classical nature 
of the initial state to a large degree.

We note that above estimates only provide an upper bound  on the
change in relative fluctuation for the states which satisfy
generic semi-classicality requirements initially. They are
consistent with the numerical simulations with more specific
states \cite{APS,APSV}. Also, one should be careful in
interpreting the results. From the above, one may be tempted to
conclude that since absolute fluctuations may change
significantly, there may be a loss of semi-classicality across the
bounce. That such a conclusion is not realized becomes clear once
one computes relative fluctuations of the observables which encode
peakedness properties of the state.

To summarize, we have shown that, within a simplified and
completely solvable model of LQC,
if the universe is semiclassical at late times, that is, if it has 
very small relative dispersions in both relevant canonically conjugate
variables, then the universe
 `before the big bang' at early times is also semiclassical. To
be precise, there is a strong bound on the possible relative
dispersion `on the other side' when the state is known to have, at
late times, small relative dispersion on canonically conjugate
variables.  This result contributes significantly to the robustness
of the sLQC scenario regarding singularity resolution \cite{ACS},
and shows that the possibilities highlighted in \cite{nature}
are not realized. A question which remains open is the variation of
higher moments quantifying more detailed properties of the state. 
Lessons learnt from the behavior of fluctuations give us valuable insights 
on their expected behavior, in particular that for realistic universes 
such variations may be bounded by very small numbers.


\noindent We thank A. Ashtekar for extensive discussions and
helpful comments and T. Pawlowski for various discussions. This
work was in part supported by CONACyT U47857-F, NSF PHY04-56913
grants and by the Eberly Research Funds of Penn State. Research at
Perimeter Institute is supported by the Government of Canada
through Industry Canada and by the Province of Ontario through the
Ministry of Research \& Innovation.

\end{document}